\documentclass[sigconf,review]{acmart}

\renewcommand\footnotetextcopyrightpermission[1]{}

\usepackage{soul}
\usepackage{amsmath,amsfonts}
\usepackage{algorithmic}
\usepackage[ruled,vlined ]{algorithm2e}
\usepackage{multirow}
\usepackage{array}
\usepackage{tabularx}
\usepackage{graphicx}
\usepackage{textcomp}
\usepackage{xcolor}
\usepackage{psfrag}
\usepackage{mathtools}
\usepackage{xspace}
\usepackage{comment}
\usepackage[normalem]{ulem}
\def\BibTeX{{\rm B\kern-.05em{\sc i\kern-.025em b}\kern-.08em
    T\kern-.1667em\lower.7ex\hbox{E}\kern-.125emX}}
    
\def\skipnoindent{\vskip0.1in\noindent}

\newcounter{challenge}

\newcounter{recommendation}

\newcounter{myposition}

\usepackage{balance}

\usepackage{listings}
\usepackage{xcolor}
 
\definecolor{codegreen}{rgb}{0,0.6,0}
\definecolor{codegray}{rgb}{0.5,0.5,0.5}
\definecolor{codepurple}{rgb}{0.58,0,0.82}
\definecolor{backcolour}{rgb}{0.95,0.95,0.92}
 
\lstdefinestyle{mystyle}{
    backgroundcolor=\color{backcolour},   
    commentstyle=\color{codegreen},
    keywordstyle=\color{magenta},
    numberstyle=\tiny\color{codegray},
    stringstyle=\color{codepurple},
    basicstyle=\ttfamily\footnotesize,
    breakatwhitespace=false,         
    breaklines=true,                 
    captionpos=b,                    
    keepspaces=true,                 
    numbers=left,                    
    numbersep=5pt,                  
    showspaces=false,                
    showstringspaces=false,
    showtabs=false,                  
    tabsize=2
}
 
\def\adityaIgnore#1{}

\begin{document}
\title{A Data-Centric Approach to Generate Invariants for a Smart Grid Using Machine Learning}

\fancyhead{}

\author{Danish Hudani}
\affiliation{%
  \institution{DHA Suffa University}
%   \streetaddress{1 Th{\o}rv{\"a}ld Circle}
%   \city{Hekla}
%   \country{Iceland}
}
\email{cs171078@dsu.edu.pk}

\author{Muhammad Haseeb}
\affiliation{%
  \institution{DHA Suffa University}
%   \streetaddress{1 Th{\o}rv{\"a}ld Circle}
%   \city{Hekla}
%   \country{Iceland}
}
\email{cs172005@dsu.edu.pk}

\author{Muhammad Taufiq}
\affiliation{%
  \institution{DHA Suffa University}
%   \streetaddress{1 Th{\o}rv{\"a}ld Circle}
%   \city{Hekla}
%   \country{Iceland}
}
\email{cs172025@dsu.edu.pk}

\author{Muhammad Azmi Umer}
\affiliation{%
  \institution{DHA Suffa University}
%   \streetaddress{1 Th{\o}rv{\"a}ld Circle}
%   \city{Hekla}
%   \country{Iceland}
}

\email{azmi.umer@dsu.edu.pk}

\author{Nandha Kumar Kandasamy}
\affiliation{%
  \institution{Singapore Institute of Technology}
%   \streetaddress{1 Th{\o}rv{\"a}ld Circle}
%   \city{Hekla}
%   \country{Iceland}
}

\email{nandha001@e.ntu.edu.sg}

%\author{Nandha Kumar Kandasamy}
%\affiliation{
 % \institution{Singapore Institute of Technology}

%}
%\email{nandha001@e.ntu.edu.sg}

\begin{abstract}
Cyber-Physical Systems (CPS) have gained popularity due to the increased requirements on their uninterrupted connectivity and process automation. Due to their connectivity over the network including intranet and internet, dependence on sensitive data, heterogeneous nature, and large-scale deployment, they are highly vulnerable to cyber-attacks. Cyber-attacks are performed by creating anomalies in the normal operation of the systems with a goal either to disrupt the operation or destroy the system completely. The study proposed here focuses on detecting those anomalies which could be the cause of cyber-attacks. This is achieved by deriving the rules that govern the physical behavior of a process within a plant. These rules are called Invariants. We have proposed a Data-Centric approach (DaC) to generate such invariants. The entire study was conducted using the operational data of a functional smart power grid which is also a living lab.
\end{abstract}

\keywords{Machine Learning, Anomaly Detection, Cyber-Physical Attack, Cyber-Physical System, Critical Infrastructure, Industrial Control System, Association Rule Mining, Electrical Power Plant}

\maketitle

\section{Introduction}
% Introduction to the CPS and the attack detection problem

Critical Infrastructure (CI) such as Electrical Power and Water Treatment Plants are integral parts of developed countries. These CIs are highly vulnerable to cyber-attacks due to the increased requirements on their uninterrupted connectivity and process automation. Therefore, it is critical to take appropriate measures to protect them from cyber-attacks. In any CI, the intrusion is detected when the plant shifts its behavior from normal state to anomalous state. Several studies have been  proposed to detect such anomalous behavior in water treatment plants \cite{adepu2016distributed, urbina2016attacking}, in power grids \cite{hashimoto2011distributed, pasqualetti2013attack}. One such approach to detect anomalous behaviour is through a set of well-defined rules. These rules are designed to detect the anomalous behavior of the CI by leveraging plant physics. These rules are often referred to as physics-based Invariants or Invariants. Rules/Invariants can be derived using different approaches like the Design-Centric approach (DeC), and the Data-Centric approach (DaC) The Design-Centric approach has its limitations \cite{UMER_Azmi_2020}, such as: 
\begin{enumerate}
    \item Many operational plant designs consist of lengthy documentation of design diagrams making it impossible to derive an efficient set of invariants.
    \item Legacy plants will not be able to use a design-centric approach because as designs evolve, the documentation becomes obsolete.
    \item DeC approach needs Experts who can decipher plant design and generate invariants of it.
    \item Confidentiality related concerns may restrict complete access to the documents for the professional generating these invariants.
\end{enumerate}

\skipnoindent In the current study, we have used DaC approach to generate invariants. We have used an unsupervised rule-based machine learning technique, Association Rule Mining (ARM) \cite{agrawal1993mining} to derive invariants from the data obtained from Electrical Power and Intelligent Control (EPIC) testbed. EPIC testbed is used as an example for operational CI. EPIC is a testbed that replicates a real electrical power grid system and is a living lab capable of supplying power to other CIs such as Water Treatment Plants \cite{adepu2018epic}. EPIC comprises four stages, namely Generation, Smart-Home, Transmission, and Micro-grid, hence can replicate the end-to-end operation of a full-scale power system. Each stage has its relays (Intelligent Electronic Devices or IEDs), local switches, a power supply unit, and PLCs (Programmable Logic Controller) connected in a fiber optic ring network. %These stages are then further connected to the master PLC. Master PLC controls the overall operations. 
The Supervisory Control and Data Acquisition (SCADA) workstation is used to monitor the entire system and also provides supervisory control \cite{adepu2018epic}. The motivation for the current study is captured in the following research questions. \skipnoindent RQ1: Can we use the DaC approach to generate invariants for a smart power grid?
RQ2: How efficient is the DaC approach in generating invariants for a smart power grid?

\skipnoindent  {\em Contribution}: A Data-Centric approach to generate invariants for a smart power grid.

% \textcolor{red}{@Azmi: It would be interesting to have some sort of metric like this, a particular unsafe state, e.g., tank overflow and number of ways (patterns mined) this can be achieved and/or number of devices to be manipulated to achieve this or number of attributes....Or support as in duration of attacks.} 

% \skipnoindent{\em Organization}: The remainder of this paper is organized as follows. Section~\ref{sec:swat} is a brief introduction to a live water treatment plant used extensively by the authors for testing anomaly detectors derived using process data. Terminology related to anomaly detectors is explained in Section~\ref{sec:anomalyDetectors}. Challenges in the design of anomaly detectors using plant data are enumerated and explained in Section~\ref{sec:challenges}. Research directions aimed at the development of methods to overcome the challenges are summarized in Section\ref{sec:futureWork}.

% \textcolor{red}{A lot of interesting stuff and paragraphs we can reuse from this paper of Ensuk \cite{Eunsuk_Sridhar_Modelbased}: following from Eunsuk paper but we can say something similar when we get the results}
%https://dspace.mit.edu/bitstream/handle/1721.1/114444/Model-based\%20security.pdf?sequence=1&isAllowed=y}

\section{EPIC Plant and dataset}

%Details of physical processes and communication networks are available in \cite{adepu2018epic}.

\subsection{EPIC Plant and Sub-processes}

% an explanation on the task of attack detection in  a water system is given here

\begin{figure*}[tbh]
    \centering
    \includegraphics[width=350pt, height=160pt]{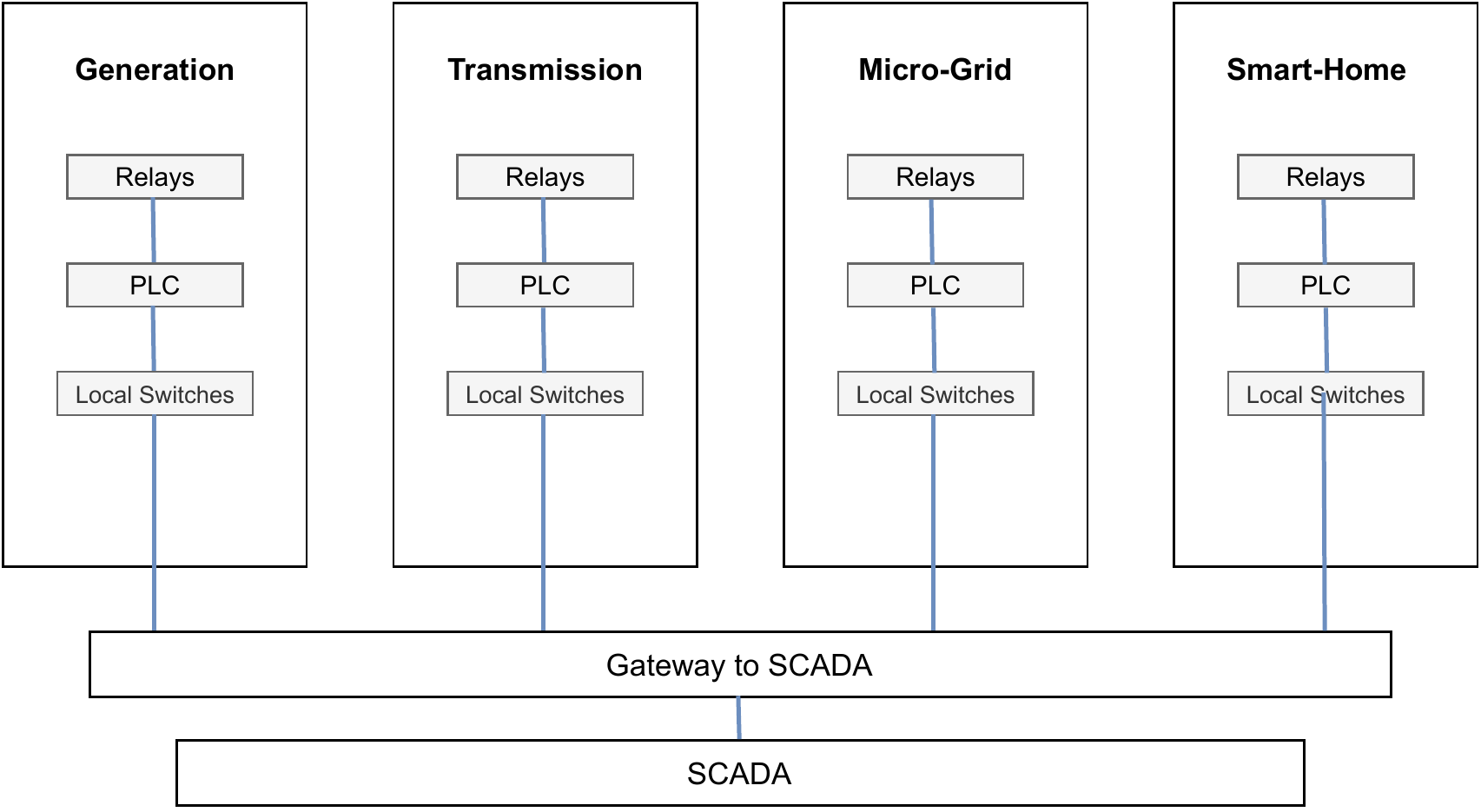}
    \caption{EPIC Control Architecture.}
    \label{fig:Epic Architecture}
\end{figure*}

Electric Power and Intelligent Control (EPIC) is a power grid testbed which consists of four stages namely Generation, Transmission, Micro-grid, and Smart Home. These stages are  connected to SCADA via a gateway (as shown in Figure \ref{fig:Epic Architecture}). SCADA provides supervisory control and is used to monitor the entire power grid. Each of the four stages of EPIC has switches, PLCs, power supply units, protection, and communication systems in a fiber optic ring network. WAGO PLCs control the opening and closing of breakers and contain the synchronization logic for the generators. High-availability Seamless Redundancy (HSR) and Media Redundancy Protocol (MRP) switches are used in the ring network for redundancy. The generation stage consists of two generators, each rated at 10kVA and providing a maximum of 20kVA. The Micro-grid stage consists of photovoltaic (PV) panels rated at 33kW and an energy storage system (ESS) with a rated power of 15kW. EPIC can be operated in two modes i.e., Micro-grid and normal. In Micro-grid mode, the generators, PV, and ESS are directly connected to the Smart-Home to represent shorter transmission lines. In normal mode, an auto-transformer (T1) is used to step up or step down the voltage at the transmission stage to the Smart-Home directly. Smart-Home consists of two load banks, 30kVA and 15kVA respectively along with a spinning load i.e., a motor to represent loads such as pumps, heat and ventilation system, etc. The motor is loaded electrically using the third generator rated at 10kVA and feeds the power back to the generation stage. Two other CIs i.e., a Water Treatment Plant and a Water Distribution Plant are also connected at the Smart-Home stage which allows these CIs to function using the power generated in EPIC. \newline
In the electric layout of EPIC \cite{adepu2018epic} the main power supply is obtained from the main grid through a circuit breaker (Main CB). It is then used to drive the variable speed drive motors (VSD1 and VSD2) referred to as M1 and M2. The generators, G1 and G2 are connected with PV panels and ESS which enables the study of the islanded or grid-connected mode \cite{adepu2018epic}. Different IEDs prefixed with their stage name protect the electrical system, e.g., GIED is an IED for protecting the Generation stage. The IEDs also function as measuring and control devices (actuators for the CBs). EPIC plant consists of four loads, referred to as water testbed and load demand. Load Demand is further sub-divided into the critical, non-critical load, and motor load. For further details on the electrical and network details the readers can refer\,\cite{ahmed2021comprehensive}%Transformer (T1) is used to control the excess load demand and/or to control the voltage deviations.

\begin{figure}[tbh]
    \centering
    \includegraphics[width=240pt, height=180pt]{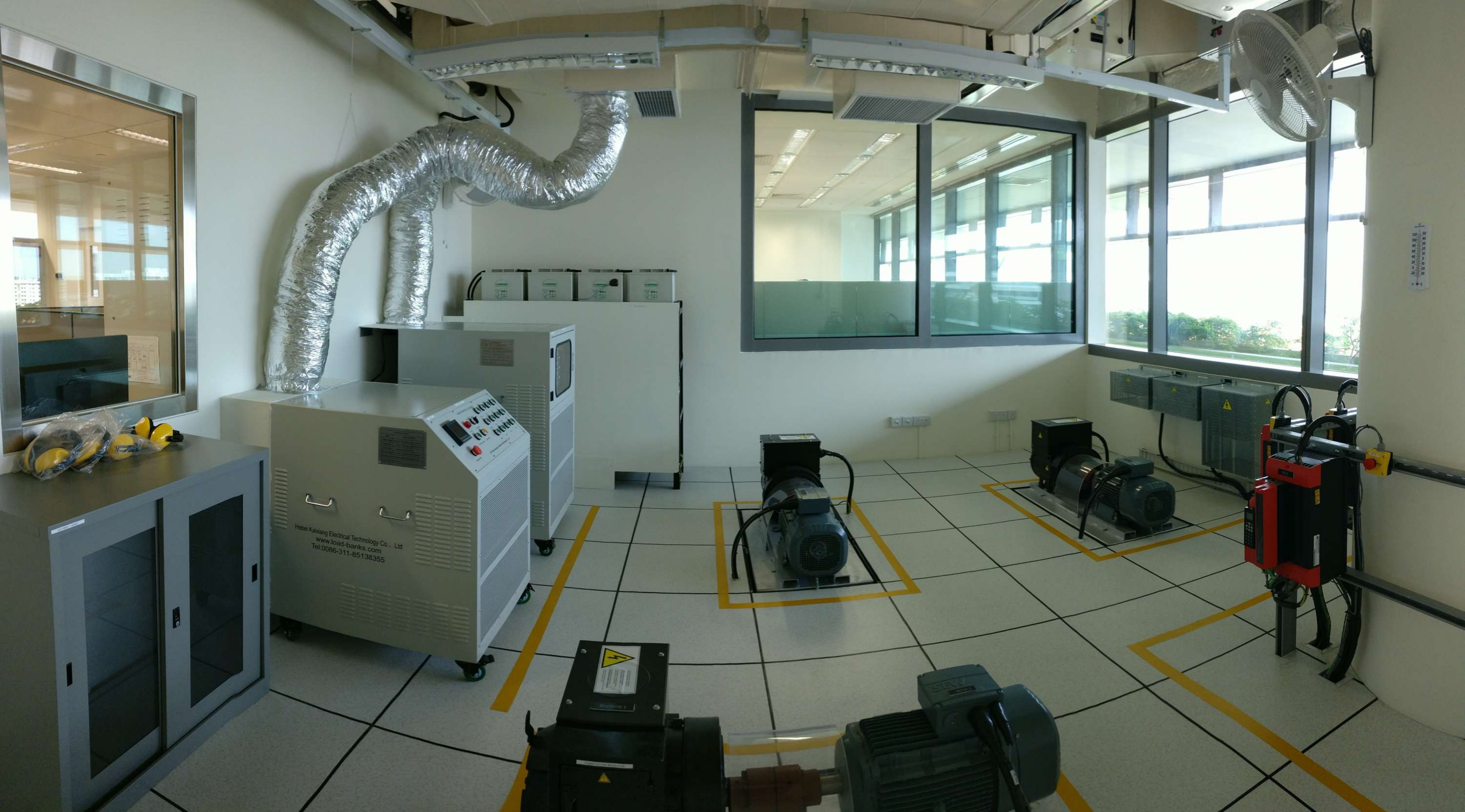}
    \caption{Case Study: EPIC Testbed}
    \label{fig:EPIC_Testbed_Pic}
\end{figure} 

%\begin{figure*}
    %\centering
    %\includegraphics[width=500, height=250]{Electrical Layout %in EPIC.pdf}
  %  \caption{Electrical Layout in EPIC.}
 %   \label{fig:Electrical Layout in EPIC}
%\end{figure*}

%Figure~\ref{fig:Electrical Layout in EPIC} shows the electrical layout of EPIC \cite{adepu2018epic}. 

\subsection{EPIC Dataset}

The authors in\,\cite{ahmed2021comprehensive} have published a comprehensive dataset for EPIC with time-stamps, the dataset was collected under multiple scenarios and  4 different scenarios of the normal operation of the EPIC testbed with a sampling rate of 1 second is considered in this paper.

\textbf{Scenario 01:} Synchronising two generators without load and angle difference from -180 to 0  and 0 to 180 degrees. The data was generated on Oct 19 2018 from 02:22 PM to 02:54 PM, having 512 data points and 292 attributes. 

\textbf{Scenario 02:} Synchronising two generators with a PV system and resistive load of 7KW for 20 minutes. The data was generated on Oct 19 2018 from 03:45 PM to 04:05 PM, having 797 data points and 292 attributes.

\textbf{Scenario 03:} Synchronising three generators with a resistive load of 4KW for around 20 minutes. The data was generated on Oct 19 2018 from 04:06 PM to 04:28 PM, having 863 data points and 292 attributes.

\textbf{Scenario 04:} Continuous supply of power to SWaT \cite{swat2016} and WADI \cite{wadi2017} testbeds of water treatment and distribution plants respectively. The data was generated on Nov 07 2018 from 02:57 PM to 03:21 PM, having 695 data points and 391 attributes.

\section{Association Rule Mining}
\label{sec:arm}

Association Rule Mining (ARM) \cite{agrawal1993mining}, also referred to as "Frequent Pattern Mining", is used to explore patterns in the dataset to find relations between different attributes. This relation is often expressed as a rule such as in \ref{eq:in}. It is a rule-based unsupervised machine learning approach in which we get antecedent followed by consequent as the output such that antecedent X implies consequent Y.

\begin{equation}
\label{eq:in}
    X   \;\Longrightarrow\; Y
\end{equation}

%\skipnoindent \textcolor{red}{ARM is an unsupervised machine learning technique where we do not have any target labels. This makes it challenging in practical applications with large datasets. As in the rule generation process, the number of rules grows exponentially concerning attributes or items in the dataset. This makes the rule generation algorithm NP-complete. To make the problem manageable, different constraints such as statistical or domain level are applied, only the rules that meet the minimum criteria of confidence, support, and lift are considered as association rules.}

\subsection{Frequent Itemsets}
ARM needs a dataset to generate frequent patterns to make the relationships discoverable across state variables. In the current study, the dataset D is the collection of sensor values from VSDs, motors, generators, and IEDs at each time interval ordered chronologically. An itemset is known as the grouping of one or more state variables, $e$ for example, is the state of VSDs, IEDs, and circuit breakers. Itemsets A which have minimum support S in D are referred to as frequent itemsets. Formally, support can be written as:

\begin{equation}
    \textstyle{S}(A)=\frac{|e\epsilon D;A\epsilon e|} {|D|}
\end{equation}

\skipnoindent Configuring low support prompts a large number of frequent itemsets which will likewise incorporate uncommon frequent itemsets, while high support will prompt few frequent itemsets and this would lead to a conservative model.

\subsection{Rules}
Frequent itemsets can also be broken down into antecedents and consequents to generate rules of the type X $\Rightarrow$ Y where X and Y are sets of items. This rule means that if transactional databases contain X, then it tends to contain Y.  Only those rules are made part of the final set of association rules that satisfies the user-defined minimum confidence threshold. Confidence C checks that how many times the rule gets true in the dataset when X has occurred. Confidence for the rule X $\Rightarrow$ Y is defined as follows.

\begin{equation}
\textstyle{C}(X \;\Rightarrow\; Y) = \frac{S(X \cup Y)} {S(X)}
\end{equation}

\skipnoindent Configuring a low value of confidence results in inaccurate rules than those generated with higher confidence values. State variables of X and Y depend on the size of frequent itemsets i.e they can have one or more state variables if the size of frequent itemsets is bigger or vice versa. 

\skipnoindent Lift is another measure that tells the likelihood of the antecedent and consequent coming together \cite{liftdef}. The lift is an implicit measure also called interest measure that how many times antecedent and consequent come together than expected if both were statistically independent. If the value of lift is less than 1 then it means that the antecedent and consequent appear less than expected. If it is greater than 1 then it means they appear more than expected, and it is equal to 1 then it means that they appear as expected. In this study, the rules mined have been filtered to the value of lift greater than or equal to 1. Lift for a rule X $\implies$ Y can be calculated using the following equation.

\begin{equation}
\textstyle{Lift}(X \;\Rightarrow\; Y) = \frac{S(X \cup Y)} {S(X).S(Y)}
\end{equation}

\section{Deriving Invariants from EPIC Dataset}
The steps involved in using the data-centric approach to mine invariants using association rule mining are described below:

%Frequent itemsets are used to generate association rules which are described in equation \ref{eq:in}. Only rules which qualify the minimum confidence threshold are considered as association rules. 

%\subsubsection{Confidence:} The rule described in equation \ref{eq:in} can be divided into two parts. One is 'X' which is placed at the left side of $\Longrightarrow$ and called as antecedent, while one is 'Y' which is placed at the right side of $\Longrightarrow$ and called as consequent. Confidence of any rule is calculated using the combined support of antecedent and consequent, and the antecedent alone.

\newcolumntype{C}[1]{>{\centering\let\newline\\\arraybackslash\hspace{0pt}}m{#1}}
\begin{table*}[]
\begin{tabular}{C{0.8cm}C{0.8cm}C{0.7cm}C{8cm}C{5cm}}
\hline
\textbf{Supp} & \textbf{Conf} & \textbf{Lift} & \textbf{Antecedent}                                                                                 & \textbf{Consequent}               \\
\hline\hline
0.959         & 1             & 1.022         & MicroGrid.Q2C.MODE\_CLOSE=False                                                                      & MicroGrid.Q2B.MODE\_CLOSE=False    \\
0.955         & 1             & 1.022         & MicroGrid.Q2B.MODE\_SYNC\_COMPLETED=False, MicroGrid.Q2C.MODE\_CLOSE=False                             & MicroGrid.Q2B.MODE\_CLOSE=False    \\
0.916         & 1             & 1.089         & Generation.Q1\_2.MODE\_CLOSE=False, Generation.Q1\_2.STATUS=CLOSE                                      & Generation.Q1\_2.STATUS\_OPEN=False \\
0.914         & 1             & 1.089         & Generation.Q1\_2.STATUS\_CLOSE=True, MicroGrid.Q2B.MODE\_SYNC\_COMPLETED=False                          & Generation.Q1\_2.STATUS=CLOSE      \\
0.914         & 1             & 1.089         & Generation.Q1\_2.MODE\_CLOSE=False, Generation.Q1\_2.MODE\_OPEN=True, Generation.Q1\_2.STATUS\_OPEN=False & Generation.Q1\_2.STATUS\_CLOSE=True\\
\hline
\end{tabular}
\caption{A sub-set of Invariants generated from Scenario 01}
\label{tab:InvSc1}
\end{table*}

\begin{table*}[]
\begin{tabular}{C{0.8cm}C{0.8cm}C{0.7cm}C{8.5cm}C{5cm}}
\hline
\textbf{Supp} & \textbf{Conf} & \textbf{Lift} & \textbf{Antecedent}                                                                                 & \textbf{Consequent}               \\
\hline\hline
0.959         & 1             & 1.034         & Generation.GIED1.Measurement.L1\_Current=0                                                                                       & MicroGrid.MIED1.Measurement.Real=0, MicroGrid.MIED2.Measurement.Apparent=0 \\
0.959         & 1             & 1.034         & Generation.GIED1.Measurement.L2\_Current=0                                                                                       & MicroGrid.MIED1.Measurement.Real=0, MicroGrid.MIED2.Measurement.Apparent=0 \\
0.956         & 1             & 1.034         & MicroGrid.MIED2.Measurement.Apparent=0                                                                                          & MicroGrid.MIED2.Measurement.Real=0                                         \\
0.955         & 1             & 1.046         & SmartHome.SIED2.Measurement.Apparent=0                                                                                          & MicroGrid.MIED2.Measurement.Apparent=0                                     \\
0.955         & 1             & 1.034         & Generation.GIED1.Measurement.L1\_Current=0, Generation.GIED1.Measurement.L2\_Current=0, Generation.GIED1.Measurement.L3\_Current=0 & MicroGrid.MIED1.Measurement.Real=0\\

\hline
\end{tabular}
\caption{A sub-set of Invariants generated from Scenario 02}
\label{tab:InvSc2}
\end{table*}

\begin{table*}[]
\begin{tabular}{C{0.8cm}C{0.8cm}C{0.7cm}C{8.5cm}C{5cm}}
\hline
\textbf{Supp} & \textbf{Conf} & \textbf{Lift} & \textbf{Antecedent}                                                                                 & \textbf{Consequent}               \\
\hline\hline
0.997         & 1             & 1.001         & MicroGrid.Q2A.MODE\_OPEN=False                                          & MicroGrid.Q2A.MODE\_CLOSE=True \\
0.997         & 1             & 1.003         & MicroGrid.Q2A.MODE\_OPEN=False                                                        & MicroGrid.Q2A.STATUS=OPEN               \\
0.997         & 1             & 1.003         & MicroGrid.Q2A.STATUS\_CLOSE=False                                                     & MicroGrid.Q2A.STATUS\_OPEN=True          \\
0.997         & 1             & 1.003         & MicroGrid.Q2A.MODE\_OPEN=False, MicroGrid.Q2A.STATUS=OPEN                             & MicroGrid.Q2A.MODE\_CLOSE=True           \\
0.995         & 1             & 1.003         & Generation.GIED1.Measurement.Power\_Factor=1, MicroGrid.Q2A.STATUS=OPEN & MicroGrid.Q2A.STATUS\_OPEN=True              
\\

\hline
\end{tabular}
\caption{A sub-set of Invariants generated from Scenario 03}
\label{tab:InvSc3}
\end{table*}

\begin{table*}[]
\begin{tabular}{C{0.8cm}C{0.8cm}C{0.7cm}C{8.5cm}C{5cm}}
\hline
\textbf{Supp} & \textbf{Conf} & \textbf{Lift} & \textbf{Antecedent}                                                                                 & \textbf{Consequent}               \\
\hline\hline
0.999         & 1             & 1.001         & MicroGrid.MAMI3.Power\_Frequency=1                                                         & MicroGrid.MAMI3.Power\_Factor=1                             \\
0.999         & 1             & 1.001         & MicroGrid.MAMI3.Voltage\_L1=1, MicroGrid.MAMI3.Voltage\_L2=1, MicroGrid.MAMI3.Voltage\_L3=1, & MicroGrid.MAMI3.Active\_Energy\_KWh=1                        \\
0.999         & 1             & 1.001         & MicroGrid.MAMI3.Power\_Frequency=1                                                         & MicroGrid.MAMI3.Active\_Energy\_KWh=1                        \\
0.999         & 1             & 1.001         & MicroGrid.MAMI3.Voltage\_L1=1, MicroGrid.MAMI3.Voltage\_L2=1, MicroGrid.MAMI3.Voltage\_L3=1, & MicroGrid.MAMI3.Power\_Frequency=1                          \\
0.993         & 1             & 1.007         & MicroGrid.MAMI2.Voltage\_L1=1, MicroGrid.MAMI3.Voltage\_L1=1                                & MicroGrid.MAMI2.Voltage\_L2=1, MicroGrid.MAMI3.Voltage\_L2=1
            
\\

\hline
\end{tabular}
\caption{A sub-set of Invariants generated from Scenario 04}
\label{tab:InvSc4}
\end{table*}

\subsection{Feature Engineering \& Selection}

\begin{figure}
    \centering
    \includegraphics[width=250pt, height=250pt]{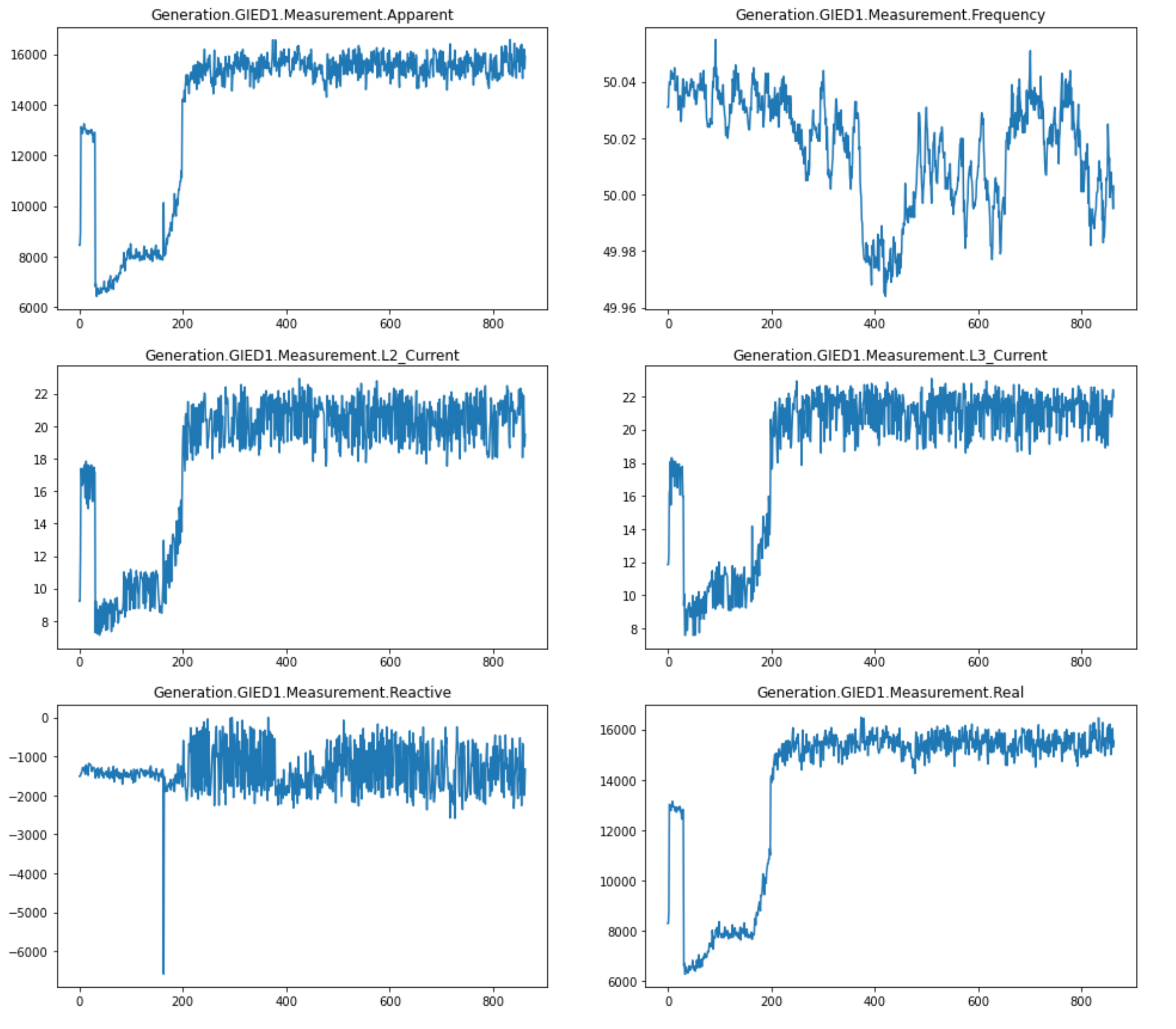}
    \caption{Attributes before Discretization}
    \label{fig:spcA}
\end{figure}

\begin{figure}
    \centering
    \includegraphics[width=250pt, height=250pt]{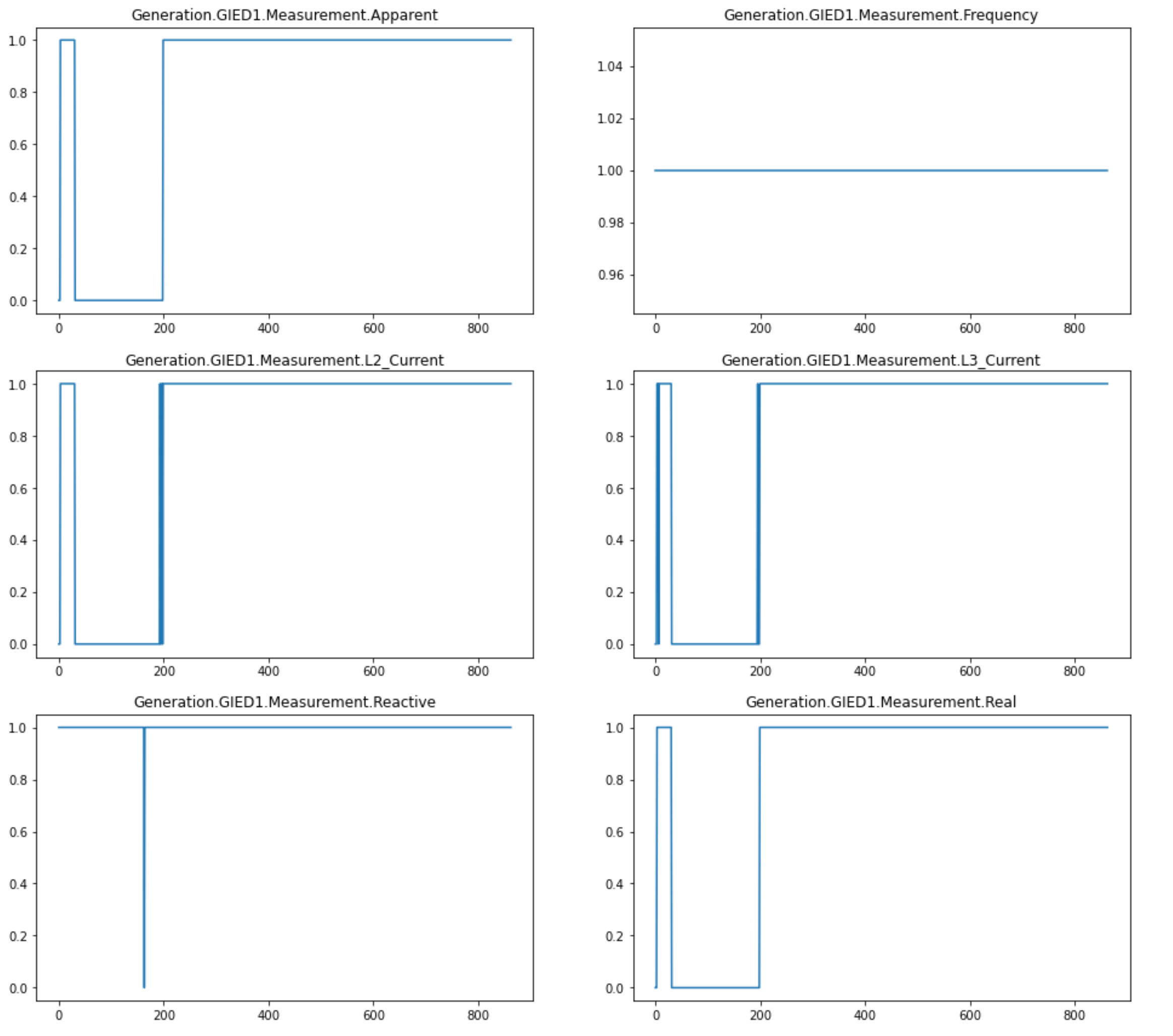}
    \caption{Attributes after Discretization}
    \label{fig:spcB}
\end{figure}

The ARM is an unsupervised machine learning technique that requires binary-valued attributes to generate finite quality rules.
In the EPIC system, we have attributes that are in different states such as continuous quantitative, discrete quantitative, and nominal qualitative. Converting these attributes into binary-valued attributes requires statistical and domain knowledge regarding electrical power systems.
Continuous quantitative attributes are converted to binary-valued attributes using interval tolerance values such as the voltage values of the attributes which should be between the intervals of either 240 ± 5\% (for phase voltages) or 420 ± 5\% (for line voltages). 
A methodical study of EPIC control strategy reveals that the states of attributes which are encoded as status codes in our dataset such as Generation.Q1.STATUS, MicroGrid Q2.STATUS, Transmission.Q3.STATUS etc. are the states of circuit breakers for each stage of the EPIC. It contains the following encoded value of 10, 01, and 11 that have been decoded to OPEN, CLOSE, and FAULT (or Trip) respectively. Some attributes are in the transition states that have been converted to their initial or final state in the system as shown in Figure \ref{fig:spcA} and \ref{fig:spcB}. For example, even though $Generation.GIED1.Measurement.Frequency$ oscillates between 49.9 to 50.1 Hz, the values can be discretized to $1$, whereas $0$ state represents values are $0$. The state and number of attributes differ in different scenarios. Many attributes in our dataset that contain a constant value throughout the operational period of the EPIC testbed have been discarded. The invariants are mined from the dynamic state of attributes that gave consequential states. Some of the attributes selected for ARM are described in Table \ref{tab:attr}.

\begin{table}
\centering
\begin{tabular}{ |p{4.1cm}|p{3.8cm}|}
 \hline
 \textbf{Attributes} & \textbf{Description}\\
 \hline
MicroGrid.MIED2.\newline Measurement.Apparent & Measures the apparent power of microgrid 02.   \\
 \hline
 MicroGrid.MIED1.Measurement.\newline Real& Measures the real power of microgrid 01. \\
 \hline
 MicroGrid.Q2B.STATUS\_CLOSE, \newline MicroGrid.Q2A.MODE\_OPEN,\newline MicroGrid.Q2C.STATUS\_OPEN & Checks the status of the circuit breakers of the microgrid.\\
 \hline
 Generation.GIED1.Measurement.\newline Reactive &   Measure the reactive power from generator 01.\\
 \hline
 SmartHome.SIED2.Measurement.\newline Apparent & Measures the apparent power of smart home 02.  \\
 \hline
 Generation.Q1\_2.MODE\_CLOSE,\newline Generation.Q1\_2.MODE\_OPEN,\newline Generation.Q1\_2.STATUS\_OPEN & Checks the status of the circuit breakers of the generator.  \\
 \hline
 MicroGrid.MAMI3.Voltage\_L1,\newline MicroGrid.MAMI3.Voltage\_L2, \newline MicroGrid.MAMI3.Voltage\_L3 & Measures the level of voltage at lines L1, L2, and L3 from the microgrid.\\
 \hline
 \end{tabular}
\caption{Some of the attributes selected for ARM}
\label{tab:attr}
\end{table}

\subsection{Invariants Generation \& Validation}
In the current study, the rules are mined using association rule mining by setting the confidence as 100\% and support as 60\% using Orange, an open-source data visualization, machine learning, and data mining tool.\newline
The rules generated under different scenarios were more than 100,000. The number of significant attributes used to derive invariants is 64, 25, 84, and 81 respectively for scenarios 01, 02, 03, and 04. Because of spatial constraint, sub-set of rules generated in scenarios 01, 02, 03, and 04 are listed in Tables \ref{tab:InvSc1}, \ref{tab:InvSc2}, \ref{tab:InvSc3}, and \ref{tab:InvSc4} respectively.

\skipnoindent The key advantage of the proposed technique in reference to DeC is the invariant generation is not linked to the effectiveness of the designer and the number of rules generated covers a wide range. Furthermore, the intrusion detection system (IDS) can be designed in such a manner to dynamically use different sub-set of invariants to improve the anomaly detection process as the adversary may not be aware of the exact set of invariants used in IDS or at least makes it harder for the adversary to design the attacks. If a fixed finite set is used it is comparatively easier for the adversary when insider information is available for carrying out the attack, such disadvantages could be minimized with the proposed approach. The validation of the rules generated could be carried out by domain experts, using different persons for sub-set of the invariants generated could drastically increase the difficulty for an adversary, as no single person will have the complete information on all sets of invariants used in the IDS.
The authors are planning to use three prong approach to validate the invariants, 1) using a random sub-set and manual validation, 2) automated validation on the EPIC plant, and 3) automated validation on the EPIC digital twin. The invariants listed in the paper were validated manually by the authors.
\section{Related Work}
Invariant generation using ARM was proposed in \cite{UMER_Azmi_2017, UMER_Azmi_2020}, which was conducted for SWaT \cite{swat2016} testbed while in the current study, invariant mining for a smart power grid was conducted which is quite different from water treatment plant. Specific attacks on EPIC process and adhoc IDS rules are presented in\,\cite{adepu2020attacks,kandasamy2019investigation}, however, rules for holistic IDS was not considered in the above papers. Similarly, ARM was also used for attacks generation in \cite{umer2021attack}. They created the attack vectors using the attacked data of SWaT. The study reported in \cite{adepu2016using} also used the process invariants to detect cyber-attacks on a water treatment system. In this study, the invariants were generated from the dynamics of the plant. They also performed physical attacks on an operational water treatment plant to evaluate their approach. Though the results of the invariant-based attack detection approach were satisfactory, attackers can still bypass the system if the parameter values are known to any attacker.

\skipnoindent Multi-level anomaly detection in Industrial Control System (ICS) using Package signatures and LSTM was proposed in \cite{feng2017multi}. Their framework was able to detect unseen attacks. It was able to deal with complicated data with hybrid features and was able to achieve high detection performance compared to other anomaly detection systems. The study in \cite{yang2006anomaly} introduced the application of two methods, SPRT and auto-associative kernel regression (AAKR) for intrusion detection in SCADA systems. This paper states the quick detection of anomalies in the system using the aforementioned methods. Different intrusion detection methods require different indicators, so monitoring potentially valuable variables is an important system requirement. The study reported in \cite{wu2019detecting} has used machine learning methods to detect cyber-attacks. They reported 96.1\% accuracy using unsupervised learning and 91.1\% accuracy using supervised learning. 

%Though numbers show that unsupervised learning has better and much more accurate results than supervised learning, but the main concern is that the labeled data of the system is not given to unsupervised learning model. Due to this reason, we have proposed ARM which is also an unsupervised machine learning technique but we have used it to derive invariants from the normal behavior of plants and to detect anomalies from it.

\section{Conclusions}
The study sought to determine how a data-centric approach can be used for the generation of invariants which would later serve as monitors for anomaly detection. While the validation phase of the data-centric approach can be made much more effective, we reach the following conclusions:
\begin{enumerate}
    \item DaC approach appears better than DeC in the sense that it considers a continuous functional mode of the plant that would tune the state of attributes. The Design-centric approach may be inefficient for tuning attributes as its states are assumed at the time when the plant is designed.
    
    \item For large-scale critical infrastructures, the generation of invariants must be automated due to a large number of sensors and components governing the functioning of the plant. For the process of automation data-centric approach is feasible.
    
    \item While the data-centric approach is more efficient in terms of cost of time we can’t deny the significance of the design-centric approach especially in the case of innovative and modern plan design.
    
    \item The number of invariants derived is in large quantity adding spatial complexity to work with them, which can be reduced by tuning the model continuously to achieve a significant set of invariants that depicts viable relations of attributes in the given system.
\end{enumerate}

\bibliographystyle{ACM-Reference-Format}
\balance
\bibliography{references.bib}

\end{document}